\documentstyle[12pt,prd,aps,epsf]{revtex}

\begin{document}
\def\abstract#1{\begin{center}{\bf ABSTRACT}\end{center}
\par #1}
\def\title#1{\begin{center}{\large {#1}}\end{center}}
\def\author#1{\begin{center}{\sc #1}\end{center}}
\def\address#1{\begin{center}{\it #1}\end{center}}

\def\pubnum{98--82}

\begin{titlepage}
\hfill
\parbox{7cm}{{YITP--\pubnum} \par Dec. 1998 \par Apr. 1999 (revised \&
  title changed)}
\parbox{6cm}{}
\par
\vspace{1.5cm}
\begin{center}
\Large
Averaging from a global point of view
\end{center}
\vskip 1cm
\author{Masayuki TANIMOTO\footnote{JSPS Research Fellow. 
Electronic mail: tanimoto@yukawa.kyoto-u.ac.jp}}

\address{Yukawa Institute for Theoretical Physics,
        Kyoto University, Kyoto 606-8502, Japan.}
\vskip 1 cm

\abstract{ We study the averaging problem from a point of view of
  variation of spatial volume $V$. We show that in the space of
  spherically symmetric dust solutions which are regular on the spatial
  manifold $S^3$ the variation $\delta V$ vanishes at the
  Friedmann-Lema\^{\i}tre-Robertson-Walker (FLRW) solution in an
  appropriate sense, which supports the validity of the FLRW solution as
  the averaged solution. We also present the second variation $\delta^2
  V$, giving the leading effect of the deviation from the FLRW
  solution. }

\end{titlepage} \addtocounter{page}{1}


\def\A{{\cal A}}
\def\B{{\cal B}}
\def\R{{\bf R}}
\def\d{{\rm d}}
\def\goes{\rightarrow}
\def\rcp#1{{1\over #1}}
\def\reff#1{(\ref{#1})}
\def\bra#1{\left[ #1 \right]}
\def\brace#1{\left\{ #1 \right\}}
\def\paren#1{\left( #1 \right)}
\def\manbo#1{}
\def\G{{\cal G}}
\def\x{{\bf x}}
\def\tc{\tau_{0{\rm c}}}

\section{Introduction}

The standard cosmology is based on the assumption that our Universe is
homogeneous and isotropic. However, the present our Universe is not
homogeneous but has clumpy structures like stars, galaxies, clusters of
galaxies, and superclusters. The recent observation \cite{COBE} of the
highly isotropic cosmic microwave background radiation is usually
regarded as an evidence of the assumption, but this is the case only up
to the stage of decoupling.  Nonetheless, one may want to think that the
homogeneous and isotropic universe model, known as the
Friedmann-Lema\^{\i}tre-Robertson-Walker (FLRW) model, reflects the
averaged nature of the true Universe over a scale larger than a
supercluster. In particular, we usually expect that the global
expansion of the Universe is well approximated by the FLRW model with
the energy distribution given by the volume average.

However, this ``averaging hypothesis'' is never justified in a trivial
way \cite{El}, since Einstein's equation is highly nonlinear. In
particular, averaging over a (spatial) volume does not commute with the
time-evolution in general, so the averaged initial data can develop in
time in a quite different way from the true data. One is therefore
required to clarify the meaning of the averaging and establish the
domain of applicability of it.

One of the earliest work connected to this problem is due to Futamase
\cite{Fu}, who built a formalism which gives the back reaction of small
scale inhomogeneities to the global expansion along the spirit of
Isaacson \cite{Isaa} using the post-Newtonian expansion. However, it is
difficult to justify the approximation used in this approach, and the
basic equations are still hard to handle. Zalaletdinov \cite{Zal}
proposed a covariant averaging formalism starting from some axioms, but
this is so complicated that it seems very difficult to draw useful
consequences. Up to now, in spite of efforts \cite{BE,RSKB,CM,IHK}
including the above, the averaging has not been well understood, because
of its complexities and conceptual difficulties.

In this paper we study the problem in a quite different way from the
conventional ones. We focus on the dynamics of the scale factor
$a(\tau)$, which is, if the space is closed \cite{note:closed},
equivalent to the dynamics of the total volume $V(\tau)$ in the sense
$V(\tau)\propto a^3(\tau)$.  We examine to what extent the FLRW solution
can be a good model judging from a point of view of the time-development
of the total volume $V(\tau)$.  Note that when given a space of
solutions spanned by some arbitrary functions of space, we can think of
$V(\tau)$ as a functional on it. By evaluating the variation of
$V(\tau)$ at the FLRW solution we may be able to obtain some information
about the quality of the FLRW solution as an ``averaged'' model. We
explicitly do this for the spherically symmetric dust case. As a result,
we find that the FLRW solution is a {\it critical point} for $V(\tau)$ in
an appropriate sense, which gives a good support for the validity of the
FLRW solution as an averaged model. We also evaluate the second
variation of $V(\tau)$, which gives the leading effect of the deviation
from the FLRW solution.

The exact solution for spatially spherically symmetric dust spacetimes
is known as the Lema\^{\i}tre-Tolman-Bondi (LTB) solution.  We add to
this solution the assumption that the spatial manifold be $S^3$.
We give a description of this subclass of the LTB solution first.

\section{The spherically symmetric dust solution on $S^3$}

The metric is written with the synchronous and
comoving coordinates as: 
\begin{equation}
  \label{eq:1}
  \d s^2=\d \tau^2-e^{\lambda(\tau,R)}\d
R^2-r^2(\tau,R)(\d\theta^2+\sin^2\theta\d\varphi^2),
\end{equation}
where $\lambda(\tau,R)$ and $r(\tau,R)$ are the functions to be
determined from the Einstein equation. The general solution to this
metric is well known as the Lema\^{\i}tre-Tolman-Bondi (LTB) solution
(see e.g. \cite{LL,Kr}), which possesses three arbitrary functions
$f(R)$, $F(R)$, and $\tau_0(R)$. (Our notation follows Ref.\cite{LL},
except for the sign of $f(R)$.) With these the function $\lambda$ is
given by $e^\lambda=r'^2/(1-f(R))$, where the dash stands for the
derivative with respect to $R$. The function $r$ is given in three
separate forms, depending upon whether the arbitrary function $f(R)$ is
negative, positive, or zero, and each solution possesses the FLRW limit
of negative ($k=-1$), positive ($k=1$), and flat ($k=0$) constant
curvature, respectively. (Such a limit is achieved when $r(\tau,R)$
separates as $r(\tau,R)=\Phi(\tau)\Psi(R)$ \cite{Kr}.)  Since our model
is spatially $S^3$, the spatial manifold does admit a constant curvature
limit and it should be positive. So, it is natural to choose the
positive sign of $f(R)$ \cite{note:fsign}, for which the the function
$r$ is given by
\begin{equation}
  \label{eq:2}
  r={F\over 2f}(1-\cos\eta),\; \tau-\tau_0(R)={F\over
  2f^{3/2}}(\eta-\sin\eta).
\end{equation}
The arbitrary function $\tau_0(R)$ is called the function of ``big bang
time'', since each Killing orbit ($R=$constant) degenerates at time
$\tau=\tau_0(R)$.

Now we have seen that the solution is parametrized by three
``arbitrary'' functions, but not all solutions are suitable for the
spatial manifold $S^3$.  The aim of the remaining part of this section
is to describe the conditions imposed on the three functions to obtain
``regular'' solutions on $S^3$. (By a regular solution we mean that it
is regular on the spatial manifold $S^3$ {\it during a finite interval
  of time} from the big bang.) We, by describing them, show that there
does exist sufficiently large set of the regular solutions in the space
of all formal solutions. This will be needed to make the formal
calculations in the next section realistic.

Let us think of the three-sphere as a sum of two balls, $S^3\simeq
D^3\cup D^3$. This is similar to the decomposition of two-sphere into
two disks, $S^2\simeq D^2\cup D^2$, achieved by cutting the two-sphere
along the equator. We can understand the spherically symmetric
three-sphere by thinking of each ball spherically symmetric. Since each
such ball has a symmetry center, the spherically symmetric three-sphere
has {\it two} symmetry centers, which are degenerate Killing orbits, as
well. We label these points as $R=0$ and $R=\pi$, so the relevant region
of the coordinate $R$ is $I\equiv [0,\pi]$. The first condition we
should impose is therefore
\begin{equation}
  \label{eq:c0}
  r=0, \; \mbox{at } R=0\mbox{ and }\pi.
\end{equation}
The function $r$ should be positive except at the boundaries.
Furthermore, we must impose some regularity conditions. An efficient way
to see this is to calculate the scalar curvature ${\cal R}$ and the
curvature scalar polynomial ${\cal R}_{abcd}{\cal R}^{abcd}$, where
${\cal R}_{abcd}$ is the curvature tensor. Using our local solution, we
find ${\cal R}=\varepsilon$ and ${\cal R}_{abcd}{\cal R}^{abcd}=
12(F/r^3)^2-8(F/r^3)\varepsilon+3\varepsilon^2$, where
$\varepsilon=F'/(r'r^2)$ is the energy density of dust. Since these
scalars must be finite, we have
\begin{equation}
  \label{eq:c1}
  {F\over r^3}<\infty, \; R\in [0,\pi]
\end{equation}
and
\begin{equation}
  \label{eq:c2}
  \varepsilon={F'\over r'r^2}<\infty, \; R\in [0,\pi].
\end{equation}
While condition \reff{eq:c1} is necessary to avoid the conical
singularity possibly generated at $R=0$ and $\pi$, condition
\reff{eq:c2} is imposed to avoid the well-known shell-crossing
singularity \cite{note:1}. Finally, we have to impose the following
coordinate condition, which is necessary for the coordinates to span the
spatial manifold $S^3$ well;
\begin{equation}
  \label{eq:c3}
  e^\lambda={r'{}^2\over 1-f}>0,\; R\in [0,\pi].
\end{equation}
Without this condition, we possibly have fictitious solutions.

To find the solution to the boundary conditions, note that there exists
the freedom of reparameterizations $R\goes \gamma(R)$.  Using this
freedom we can fix the leading power of $r$ at the boundaries to the
unity, i.e., we can make
\begin{equation}
  \label{eq:r}
  r\propto \left\{
\begin{array}{ll}
R \quad & (R\goes 0) \\
\pi-R \quad & (R\goes \pi)
\end{array}
\right. .
\end{equation}
Near the big bang singularity $(\eta\goes 0)$, from Eq.\reff{eq:2} we
have $r\sim (F/(4f))\eta^2$, and $\tau-\tau_0(R)\sim
(F/(12f^{3/2}))\eta^3$, so we find $r\sim
(9/4)^{1/3}F^{1/3}(\tau-\tau_0(R))^{2/3}$. Taking the condition
\reff{eq:r} into account, we have
\begin{equation}
  \label{eq:4}
  F(R)\propto \left\{
\begin{array}{ll}
R^3 \quad & (R\goes 0) \\
(\pi-R)^3 \quad & (R\goes \pi)
\end{array}
\right. .
\end{equation}

Let $\sigma$ be a real number, and suppose $\eta\propto R^\sigma$
($R\goes 0$) for an arbitrary fixed $\tau$. From Eq.\reff{eq:2}, we find
that if $\sigma < 0$, the function $r$ would oscillate heavily between
positive values and zero near the boundary. So, this case is unsuitable
for the condition.  On the other hand, if $\sigma\geq 0$, the boundary
condition \reff{eq:c0} is satisfied and the behavior of $f(R)$ near the
boundaries can be determined as
\begin{equation}
  \label{eq:3}
  f(R)\propto \left\{
\begin{array}{ll}
R^{\alpha_1} \quad & (R\goes 0) \\
(\pi-R)^{\alpha_2} \quad & (R\goes \pi)
\end{array}
\right. ,
\end{equation}
where $\alpha_1,\alpha_2\geq 2$. The condition \reff{eq:c1} is now trivially
satisfied.

\begin{figure}[btp]
  \begin{center}
    \leavevmode
\epsfysize=4cm
\epsfbox{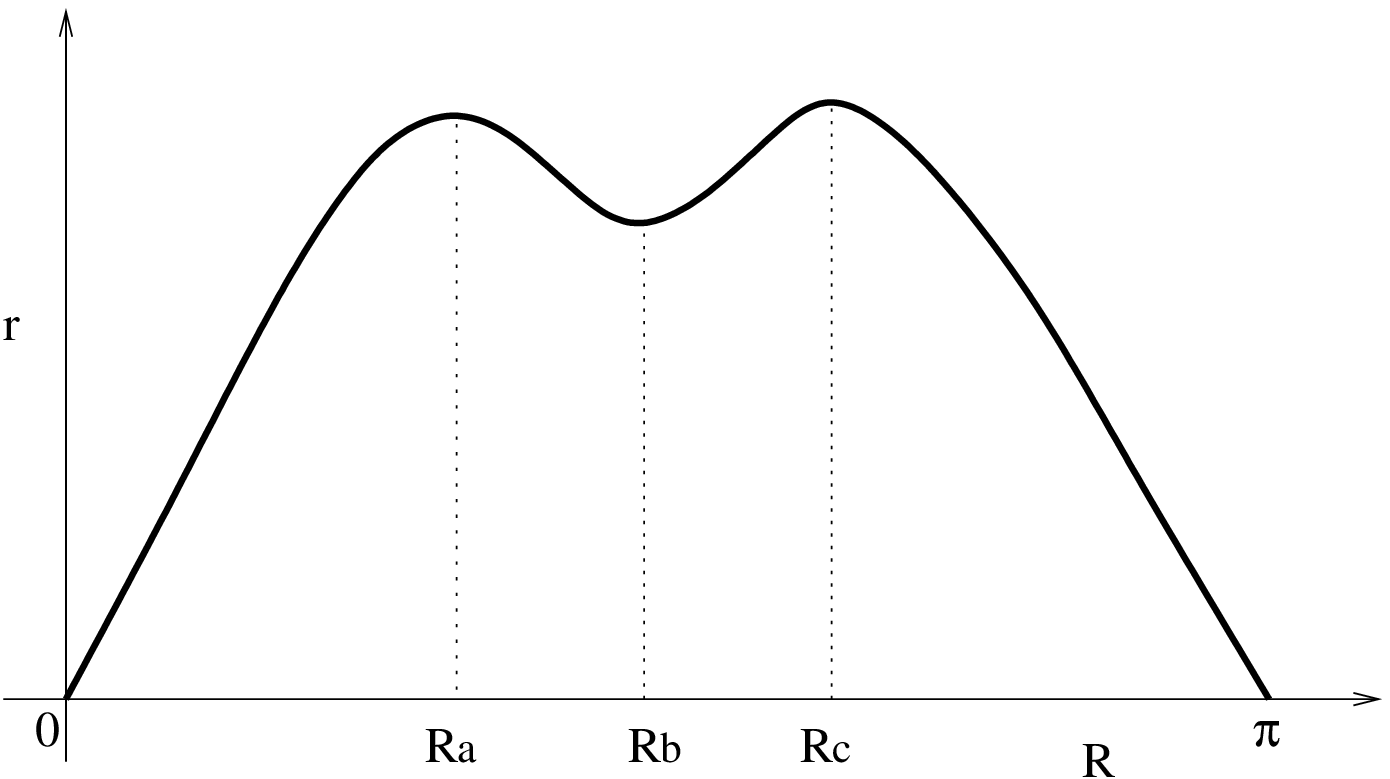}
  \end{center}
\caption{ A possible profile of $r$. There are three extremal points
  $R=R_a,R_b,R_c$ in this example.}
\label{fig:f1}
\end{figure}

\begin{figure}[btp]
  \begin{center}
    \leavevmode
\epsfysize=4cm
\epsfbox{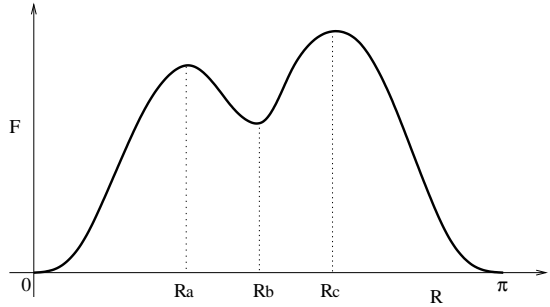}
  \end{center}
\caption{ A possible profile of $F$. The extremal points decide those
  for $r$.}
\label{fig:f2}
\end{figure}

\begin{figure}[btp]
  \begin{center}
    \leavevmode
\epsfysize=4cm
\epsfbox{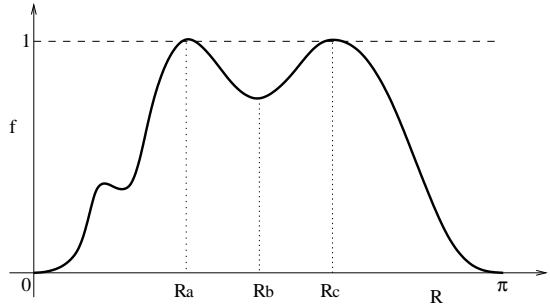}
  \end{center}
\caption{ A possible profile of $f$. At the extremal points of $F$,
  $f$ should also be extremal, and at the maximal points among them $f$
  should take value 1.}
\label{fig:f3}
\end{figure}

\begin{figure}[btp]
  \begin{center}
    \leavevmode
\epsfysize=4cm
\epsfbox{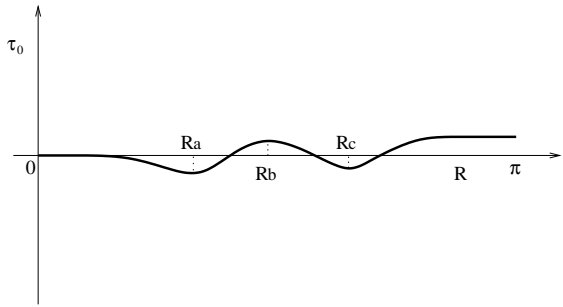}
  \end{center}
\caption{ A possible profile of $\tau_0$. The sign of the derivative
  should be opposite to that of $F$ or equal to zero.}
\label{fig:f4}
\end{figure}

Next, consider the regularity conditions \reff{eq:c2} and \reff{eq:c3}.
Note that the function $r$ on $I$ for a fixed $\tau$ has at least one
extremum (FIG.\ref{fig:f1}), since $r>0$ on the interior of $I$ and
$r=0$ at the boundaries.  From the condition \reff{eq:c2}, we find that
the function $F'$ should vanish where $r'$ vanishes
(FIG.\ref{fig:f2}). Since $F(R)$ is independent of $\tau$, $r'$ should
also vanish on $I$ independently from $\tau$. Otherwise, the regularity
would break instantaneously. A straightforward calculation gives
\begin{equation}
  \label{eq:rd}
  r'=\rcp{1-\cos\eta}\bra{\frac{F'}{2f}\A(\eta)+
    \frac{Ff'}{2f^2}\B(\eta)-f^{1/2}\tau_0'\sin\eta},
\end{equation}
where $\A(\eta)\equiv (1-\cos\eta)^2-\sin\eta(\eta-\sin\eta)$, and
$\B(\eta)\equiv -(1-\cos\eta)^2+\frac32\sin\eta(\eta-\sin\eta)$.  Since
$r'$ is a homogeneous linear combination of $f'$, $F'$, and $\tau_0'$
with distinct coefficients as functions of $\tau$, the only possible
points on $I$ for which $r'$ vanish independently from $\tau$ are the
points where $F'$, $f'$ and $\tau_0'$ vanish simultaneously.  In
particular, $r'$ and $F'$ should have the same sign everywhere on $I$
since otherwise the energy density $\varepsilon$ would become negative
in some regions. We exclude such a case for physical reason. On the
other hand, the sign of $f'$ is rather arbitrary, except that $f(R)$
should be maximal and take value $1$ where $F'$ vanishes
(FIG.\ref{fig:f3}). Taking value $1$ is a consequence of the condition
\reff{eq:c3}.  The sign of $\tau_0'$ should be opposite to $F'$ or zero
(FIG.\ref{fig:f4}). This is a consequence of the condition that $r'$
have the same sign as $F'$ at least for a finite interval of time from
the big bang ($\eta=0$). In fact, the leading powers of $\A(\eta)$,
$\B(\eta)$, and $\sin\eta$ in Eq.\reff{eq:rd} are, respectively,
$\rcp{12}\eta^4$, $-\rcp{80}\eta^6$, and $\eta$, so the term
proportional to $\tau_0'$ dominates near the big bang singularity if
$\tau_0'$ does not vanish.  It is clear that if $\tau_0'$ had the same
sign as $F'$, $r'$ would have the opposite sign when $\eta\goes 0$.

Now, we have spelled out all the conditions for $F(R)\geq0$,
$f(R)\geq0$, and $\tau_0(R)$. They are Eqs.\reff{eq:4} and \reff{eq:3},
that $f(R)$ should take maximum value $1$ where $F'$ vanishes, and that
the sign of $\tau_0'$ should be opposite to $F'$ or zero.  It is nice to
keep in mind that the profile of $F(R)$ decides that of $r$ if
the solution is regular.

\section{Variation of $V$ and averaging}

Given an inhomogeneous solution, in what way can we measure the
resemblance between the solution and the FLRW solution? We note that the
only dynamical content of the FLRW model is the scale factor $a(\tau)$,
which can be regarded as the (cubic root of the) total volume $V(\tau)$
if the spatial manifold is closed \cite{note:closed}. Since $V(\tau)$ is
always well defined for any spatially closed inhomogeneous spacetime
\cite{note:4} there is a large amount of theoretical advantage to
utilize it. Note that $V$ can be considered as a functional
$V[\phi_1(\x),\cdots \phi_n(\x)]$ on a space of solutions, where
$\phi_1(\x),\cdots \phi_n(\x)$ are the arbitrary functions of space
which span the space of solutions.  For example, in the spherically
symmetric dust case, $n=3$ and we can put
$(\phi_1(\x),\phi_2(\x),\phi_3(\x))\equiv (F(R),f(R),\tau_0(R))$. If the
inhomogeneous solution is not too far from the FLRW solution, we can
compare the two solutions by evaluating the variation $\delta V(\tau)$
at the FLRW solution. If this function $\delta V(\tau)$ vanishes for all
$\tau$, we may say that the two solutions are dynamically close to each
other and the ``averaged'' solution of the inhomogeneous solution
corresponds to the FLRW solution.

Below, we apply the above idea to the spherically symmetric dust
solution on $S^3$. Our space $P$ of solutions are assumed to be spanned
by the regular solutions on $S^3$, and the variations are taken in
it. More specifically, $P$ is a subspace of the larger space $P^*$ which
is spanned by all possible configurations of three smooth functions
$(F(R),f(R),\tau_0(R))$, and defined by subtracting from $P^*$ the
configurations which correspond to irregular solutions. The space $P$
contains the FLRW solution, since this solution is regular.

The total volume for the metric \reff{eq:1} is
\begin{equation}
  \label{eq:3-1}
  V(\tau) = 4\pi\int_0^\pi e^{\lambda/2}r^2\d R,
\end{equation}
where $e^{\lambda}=r'{}^2/(1-f(R))$ and $r$ is given by
Eq.\reff{eq:2}. For later use, we also define the total energy $E$:
\begin{equation}
  \label{eq:3-5}
  E\equiv \int_{S^3}\varepsilon\d V=4\pi\int_0^\pi
  \frac{|F'|}{\sqrt{1-f}}\d R,
\end{equation}
which is a conserved quantity. Moreover, for convenience we change the
parameterization $(F(R),f(R),\tau_0(R))$ to $(A(R),f(R),\tau_0(R))$
defined by $ A(R)\equiv F/(2f^{3/2})$. The standard FLRW limit is
achieved in this parameterization when
\begin{equation}
  \label{eq:3-3}
  (A(R),f(R),\tau_0(R))=(a_0,\sin^2 R,\tc),
\end{equation}
where $a_0$ is a positive constant parameter, and $\tc$ is another
constant parameter. Since we can always choose $\tc=0$ by the coordinate
transformation $\tau\goes\tau+\tc$, parameter $\tc$ is redundant as far
as the FLRW solution is concerned, but we will find that in our wider
context it is useful not to fix this parameter. At the limit
\reff{eq:3-3}, $V$ is given by
\begin{equation}
  \label{eq:EV0}
  V_0(\tau)=2\pi^2a_0{}^3(1-\cos\eta)^3, \; \tau-\tc=a_0(\eta-\sin\eta).
\end{equation}

We vary the volume $V$ with respect to $A(R)$, $f(R)$, and $\tau_0(R)$,
and evaluate at the FLRW solution \reff{eq:3-3}. The formal result is
\begin{eqnarray}
  \label{eq:3-3.5}
  \delta V(\tau)= && 4\pi a_0^2 (1-\cos\eta)\A(\eta)\int_0^\pi \paren{
    3\delta A+\tan R\delta A'}\sin^2 R\,\d R \nonumber \\ 
  && +2\pi a_0^3
  (1-\cos\eta)^3\int_0^\pi \paren{\tan R\, \delta f}' \d R \nonumber \\ 
  && -4\pi a_0^2 (1-\cos\eta)\sin\eta\int_0^\pi \paren{3\delta \tau_0+\tan
    R\delta \tau_0'}\sin^2 R\,\d R. 
\end{eqnarray}
We may expect that the second term vanishes if we take into account the
boundary conditions in the previous section, but we have to check the
continuity of the function $\tan R\delta f$ at $R=\pi/2$ to do this. The
zeroth variation of $f(R)$ is $\sin^2 R$ and the maximal value should
always be 1 as we explained in the previous section. This means that at
the neighborhood of $R=\pi/2$, the significant (or ``dangerous'')
variation of $f(R)$ should always be approximated by $\sin^2 (R+a)$,
where $a$ is a parameter. Differentiating this with respect to $a$ and
putting $a=0$, we find that $\delta f$ is approximated by $2\sin R\cos
R\, \d a$, so that $\tan R\,\delta f$ is approximated by $2\sin^2 R\, \d
a$, which is continuous at $R=\pi/2$. We can hence safely omit the term.
(In contrast to this, we cannot apply ``integrations by parts'' to the
first and the third terms due to the discontinuities of $\sin^2 R\tan
R\delta A$ and the similar term for $\delta\tau_0$.)  Thus we have
\begin{eqnarray}
  \label{eq:3-4}
  \delta V(\tau)= && 4\pi a_0^2 (1-\cos\eta)\A(\eta)\int_0^\pi \paren{
    3\delta A+\tan R\delta A'}\sin^2 R\,\d R \nonumber \\ 
  && -4\pi a_0^2 (1-\cos\eta)\sin\eta\int_0^\pi \paren{3\delta \tau_0+\tan
    R\delta \tau_0'}\sin^2 R\,\d R. 
\end{eqnarray}

Note that the function $\eta$ depends only on $\tau$ when it concerns
the FLRW solution, so the functions of $\eta$ in the above expressions
have been factored out of the integrals. Because of this feature $\delta
V(\tau)$ becomes constantly zero, if (and only if) the variation of
$A(R)$ and $\tau_0(R)$ are taken so that the integrals in
Eq.\reff{eq:3-4} vanish. This means that the FLRW solution is a critical
point for the functional $V(\tau)$, if we restrict the directions of the
variation in that way.

The meaning of the particular directions becomes clear if we consider
the variation of the total energy $E$, which coincides with the first
term of the rhs of Eq.\reff{eq:3-4} up to prefactor:
\begin{equation}
  \label{eq:3-6}
  \delta E = 8\pi \int_0^\pi\paren{3\delta A+\tan R\delta A'}\sin^2
  R\,\d R.
\end{equation}
(We have dropped the zero term $12\pi a_0\int_0^\pi (\tan R\delta f)'\d
R$.) One may notice that the second term of the rhs of Eq.\reff{eq:3-4}
is given similarly by the variation of
\begin{equation}
  \label{eq:defC}
  C\equiv -4\pi\int_0^\pi \frac{|(f^{3/2}\tau_0)'|}{\sqrt{1-f}}\d R,
\end{equation}
which has been defined by replacing $A(R)$ by $\tau_0(R)$ (and
multiplying the factor $-1/2$) in the definition of $E$. In fact, we
obtain
\begin{equation}
  \label{eq:3-8}
  \delta C= -4\pi\int_0^\pi\paren{3\delta \tau_0+\tan R\delta \tau_0'}\sin^2
  R\,\d R,
\end{equation}
so that we can write the formula \reff{eq:3-4} as
\begin{equation}
  \label{eq:3-7}
  \delta V(\tau)=a_0^2(1-\cos\eta)\paren{\A(\eta)\,\frac{\delta E}{2}+
  \sin\eta\,\delta C}.
\end{equation}

Here, note that the space of solutions $P$ is foliated by the surface
$P_{E,C}$ of constant $E$ and $C$, while the FLRW solution $O$ is a two
dimensional subset in $P$ spanned by the two parameters $a_0$ and $\tc$.
We write the FLRW solution with fixed values of $a_0$ and $\tc$ as
$O_{a_0,\tc}\in O$. We can easily find that there is a unique FLRW
solution $O_{a_0,\tc}$ in each $P_{E,C}$, and we can define a map $Av:
P\goes O$ by this correspondence:
\begin{equation}
  \label{eq:Av}
  Av: P_{E,C}\goes O_{a_0,\tc}.
\end{equation}
The significance of Eq.\reff{eq:3-7} is that the FLRW solution
$O_{a_0,\tc}$ that best matches a given inhomogeneous solution $p\in P$
is given by $Av(p)$, the FLRW solution with the same energy $E$ and the
same ``$C$''. That is, since in each surface $P_{E,C}$ the FLRW solution
$Av(P_{E,C})$ is the critical point for the volume $V(\tau)$, all the
inhomogeneous solutions which are sufficiently close to $Av(P_{E,C})$
virtually manifest the same dynamical evolutions of volume as that of
$Av(P_{E,C})$. This is our main result. We will call $Av$ the {\it
  averaging map}.

The key relation \reff{eq:3-7} seems very reasonable (though it is not
trivial) if we note the following fact about the FLRW solution. That is,
if we vary the two FLRW parameters $a_0$ and $\tc$ the total volume
$V_0(\tau)$ for the FLRW solution will suffer a certain change. We can
easily estimate it by differentiating $V_0(\tau)$ with respect to the
two parameters. (See Eq.\reff{eq:EV0}.) The result is
\begin{equation}
  \label{eq:dV0}
  \d V_0(\tau) = 6\pi^2 a_0^2(1-\cos\eta)
  \paren{2\A(\eta) \d a_0-\sin\eta\d\tc}.
\end{equation}
Moreover, using
\begin{equation}
  \label{eq:EC0}
  E_0=12\pi^2a_0,\quad C_0= - 6\pi^2\tc
\end{equation}
for the value of $E$ and
$C$ at the FLRW limit \reff{eq:3-3}, we obtain exactly the same formula
as Eq.\reff{eq:3-7}:
\begin{equation}
  \label{eq:dV02}
  \d V_0(\tau) = a_0^2(1-\cos\eta)\paren{\A(\eta) 
    \frac{\d E_0}{2}+\sin\eta\, \d C_0}.
\end{equation}
In this sense the functionals $E$ and $C$ are ``inhomogeneous
generalizations'' of the FLRW parameters $a_0$ and $\tc$. Relations
\reff{eq:EC0} also imply
\begin{equation}
  \label{eq:Av2}
  Av(P_{E,C})=O_{a_0=\frac{E}{12\pi^2},\tc=\frac{-C}{6\pi^2}}.
\end{equation}

Recall that the parameter $\tc$ contained in the FLRW solution is merely
a kind of gauge freedom, and therefore we could fix it like
$\tc=0$. This seems to mean that we can restrict the space of solution
$P$ to the slice $C=0$ by gauge fixing. This is in fact the case.  Note
that the coordinate transformation $\tau\goes \tau-c$, where $c$ is a
constant, induces the transformation $\tau_0(R)\goes \tau_0(R)+c$. Hence
the space of solutions $P$ contains gauge freedom of this type. Since
the above transformation for $\tau_0(R)$ shifts the value of $C$ by
$4\pi c\int_0^\pi \frac{|f^{3/2}|}{\sqrt{1-f}}\d R$, every surface of
constant $C$ corresponds to the same set of spacetime solutions, and
this shows the claim. Note that the gauge freedom we consider is the
freedom of choosing the origin of the time coordinate. The
restriction $C=0$ therefore provides us a natural way of choosing the
origin of the time coordinate for every distinct inhomogeneous
solutions, especially with non-simultaneous big bangs, viewing from the
comparisons of $V(\tau)$.

We should comment on a specialty of the FLRW solution. It was natural to
naively expect that the FLRW solution describes the averaged spacetime
of an inhomogeneous spacetime that is homogeneous and isotropic over a
large scale. We have seen that this is in fact the case for the
spherically symmetric case in the sense $\delta V(\tau)=0$ at the FLRW
solution. One might similarly expect that a spacetime that is smooth
over a large scale but fluctuates over small scales could be
approximated by a smoothed solution, or one might expect that a
homogeneous but anisotropic (i.e., Bianchi or Kantowski-Sachs-Nariai
type \cite{KTH}) spacetime solution could be served as a good model for
an inhomogeneous spacetime that is homogeneous over a large
scale. However, at least for the spherically symmetric case only the
FLRW solution is the good averaged or smoothed spacetime model, since
there does not exist a point where the function $\eta(\tau,R)$ becomes a
function of only time $\tau$ in $P$, except at the FLRW solution, i.e.,
$\delta V(\tau)=0$ holds only at the FLRW solution.

The accuracy of the FLRW solution as the averaged one can be estimated
from the second variation $\delta^2V$. That is, we first parameterize
the arbitrary functions with one parameter $\epsilon$ so that
$\epsilon=0$ corresponds to the FLRW solution, and expand them in power
of $\epsilon$: $A(R;\epsilon)=a_0+\epsilon \delta A(R)+(1/2)\epsilon^2
\delta^2 A(R)+\cdots$, and similarly for $f(R;\epsilon)$ and
$\tau_0(R;\epsilon)$. Then, we expand
$V[A(R;\epsilon),f(R;\epsilon),\tau_0(R;\epsilon)]$ in accordance with
these expansions;
\begin{equation}
  \label{eq:20}
  V=V_0+ \epsilon \delta V+\frac12 \epsilon^2\delta^2
  V+\cdots.
\end{equation}
If the first order term vanishes as in our case, the second order term
$\frac12 \epsilon^2\delta^2V$ gives the leading term for the
deviation from the FLRW solution, i.e., the accuracy is given by
$v(\tau)\equiv (V-V_0)/V_0\simeq \frac12 \epsilon^2\delta^2V/V_0$.

After a lengthy calculation and a similar consideration \cite{note:omit}
as for $\delta V$, we obtain
\begin{eqnarray}
  \label{eq:secV}
  \delta^2 V &=& a_0^2(1-\cos\eta)\paren{\A(\eta)\, \frac{\delta^2
      E}{2}+ \sin\eta\, \delta^2 C} \nonumber \\ && +12\pi a_0\bigg(
  \G(\eta)J[\delta A,\delta A] +(1+2\cos\eta)J[\delta \tau_0,\delta
  \tau_0] \nonumber \\ && \hspace{7em} -
  2(3\sin\eta-\eta(1+2\cos\eta))J[\delta A,\delta\tau_0]\bigg),
\end{eqnarray}
where $\delta^2 E$ and $\delta^2 C$ are the second variations of,
respectively, the total energy $E$ and the conserved quantity $C$:
\begin{eqnarray}
  \label{eq:secE}
  \delta^2 E &=& 24\pi\int_0^\pi\paren{\delta^2 A+\rcp3\tan
  R\delta^2 A'}\sin^2R\,\d R \nonumber \\
& &  +24\pi\int_0^\pi \paren{\frac{\delta A\delta f}{\cos^2R}+\tan
  R\delta A\delta f'+\frac{\sin R}{\cos^3 R}\paren{1-\frac23\sin^2
  R}\delta f\delta A'}\d R, 
\end{eqnarray}
\begin{eqnarray}
  \label{eq:secC}
  \delta^2 C &=& -12\pi\int_0^\pi\paren{\delta^2 \tau_0+\rcp3\tan
  R\delta^2 \tau_0'}\sin^2R\,\d R \nonumber \\
& &  -12\pi\int_0^\pi \paren{\frac{\delta \tau_0\delta f}{\cos^2R}+\tan
  R\delta \tau_0\delta f'+\frac{\sin R}{\cos^3 R}\paren{1-\frac23\sin^2
  R}\delta f\delta \tau_0'}\d R, 
\end{eqnarray}
and we have defined
\begin{equation}
  \label{eq:J}
  J[\cdot,*]\equiv \int_0^\pi\paren{(\cdot)(*)+\rcp3\tan
  R\paren{(\cdot)'(*)+(\cdot)(*)'}}\sin^2 R\,\d R, 
\end{equation}
and $\G(\eta)\equiv
7-8\cos\eta+\cos^2\eta-6\eta\sin\eta+\eta^2(1+2\cos\eta)$. (If we were
allowed to apply ``integrations by parts'', the above formulae would
become much simpler, but because of the same reason as
in the calculation of $\delta V$, we could not do so. A formal application
would cause a divergence of the result.)

Since our variations are taken in the surface $P_{E,C}$ of constant $E$
and $C$ the variations of any order of $E$ and $C$ vanish, in
particular, $\delta^2 E=0$ and $\delta^2 C=0$. Hence we obtain
\begin{eqnarray}
  \label{eq:secV2}
  \delta^2 V(\tau) &=& 
  12\pi a_0\bigg( \G(\eta)J[\delta A,\delta A] +
  (1+2\cos\eta)J[\delta \tau_0,\delta \tau_0]
    \nonumber \\ && 
    \hspace{7em} 
    -2(3\sin\eta-\eta(1+2\cos\eta))J[\delta A,\delta\tau_0]\bigg).
\end{eqnarray}
This depends only upon the first variations of the functions $A(R)$
and $\tau_0(R)$.

\begin{figure}[btp]
  \begin{center}
    \leavevmode
\epsfysize=4cm
\epsfbox{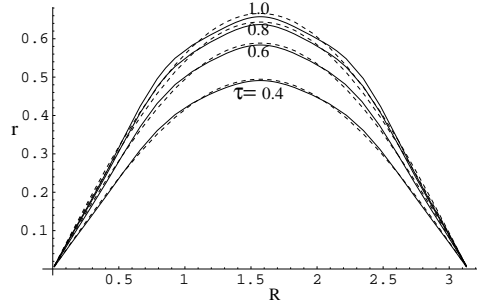}
  \end{center}
\caption{ Profiles of $r$ for the example.}
\label{fig:1}
\end{figure}

\begin{figure}[btp]
  \begin{center}
    \leavevmode
\epsfysize=4cm
\epsfbox{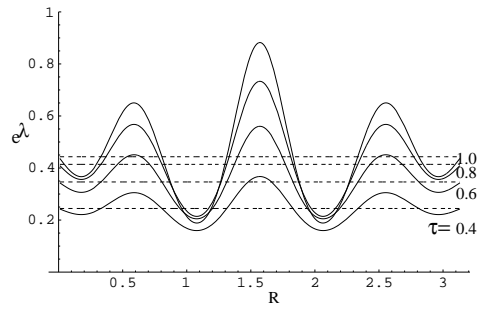}
  \end{center}
\caption{ Profiles of $e^\lambda$ for the example.}
\label{fig:2}
\end{figure}

\begin{figure}[btp]
  \begin{center}
    \leavevmode
\epsfysize=4cm
\epsfbox{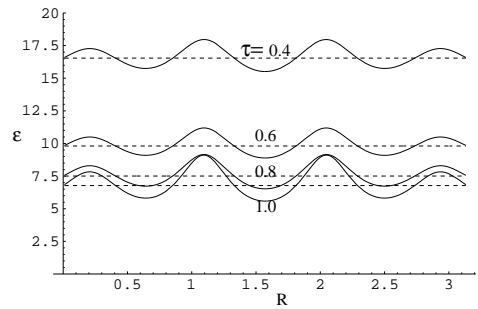}
  \end{center}
\caption{ Profiles of $\varepsilon$ for the example.}
\label{fig:3}
\end{figure}

\begin{figure}[btp]
  \begin{center}
    \leavevmode
\epsfysize=4cm
\epsfbox{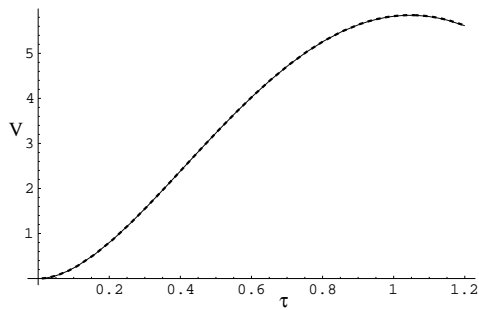}
  \end{center}
\caption{ The time-development of $V(\tau)$ for the example.}
\label{fig:4}
\end{figure}

\begin{figure}[btp]
  \begin{center}
    \leavevmode
\epsfysize=5cm
\epsfbox{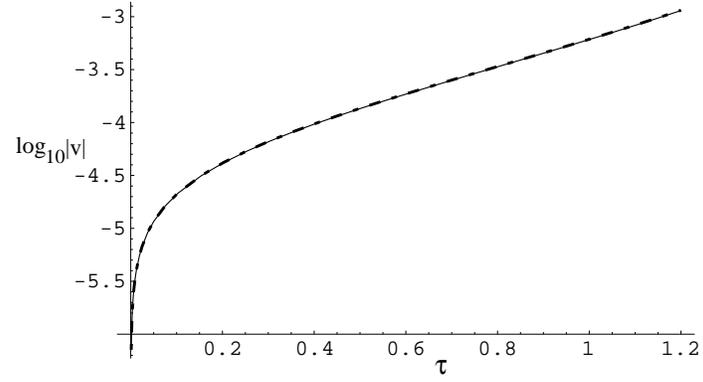}
  \end{center}
\caption{ Accuracy of $V_0(\tau)$ as the averaged solution ---
  evaluated from the second variation
  $\log_{10}\frac{\epsilon^2}2|\delta^2 V|/2V_0$ for the solid curve and
  from the exact one $\log_{10}|V-V_0|/V_0$ for the dashed curve.}
\label{fig:5}
\end{figure}

We close this section by giving an explicit example. The three functions
are $f(R)=\sin^2 R$, $A(R)=1/3+(1/900)(\sin 3R - 12\sin 5R -9\sin 7R)$,
and $\tau_0(R)=0$. (This example has been made by taking $f(R)=\sin^2 R$
and putting
\begin{equation}
  \label{eq:ex1}
  F(R)=\int_0^R f'(x)f^{\frac 12}(x)\psi(x)\d x.
\end{equation}
We can check that all the regularity conditions for $F(R)$ are satisfied
if $\psi(R)$ is a positive function on $[0,\pi]$ such that it makes
$F(\pi)=0$. In particular, if we choose $\psi(R)=1+(1/10)((1/3)\sin
5R-\sin 7R)$, we have $F(R)=\sin^3R(2/3+(1/450)(\sin 3R-12\sin 5R-9\sin
7R))$, and the $A(R)$ presented above.) The corresponding FLRW solution
as the averaged one is given by $a_0=1/3$ and $\tc=0$. FIGs.\ref{fig:1}
to \ref{fig:3} are, respectively, profiles of the metric components $r$
and $e^\lambda$, and those of the energy density
$\varepsilon$. FIG.\ref{fig:4} shows the time-development of the total
volume $V$. In each figure, the profiles of the corresponding FLRW
solution are also depicted with dashed curves.  FIG.\ref{fig:5} shows
$\log_{10}\frac{\epsilon^2}2|\delta^2 V|/V_0$, which is the estimation
of the accuracy $\log_{10} |v|$ by the second variation $\delta^2 V$,
and the dashed curve shows the exact one $\log_{10}|V-V_0|/V_0$. The
second variation $\delta^2 V$ is evaluated by putting $\epsilon \delta
A=(1/900)(\sin 3R - 12\sin 5R - 9\sin 7R)$, so the integral
$\epsilon^2J[\delta A,\delta A]=-11\pi/324000\simeq - 10^{-4}$.  We can
see that the accuracy $v$ is in this example better than $10^{-3}$
throughout the expansion phase, though the energy fluctuation becomes
larger than $10^{-1}$.

\section{Conclusions}

We have seen that the FLRW solution is the critical point for the volume
$V(\tau)$ in the space of spherically symmetric dust solutions on $S^3$.
In accordance with the fact that the FLRW solution contains two
parameters (,though one of which is redundant in a sense), we found that
there is a natural foliation in the space of solutions defined by
constant energy $E$ and another quantity ``$C$''. The exact statement of
our result is that in each leaf of the foliation there exist a unique
FLRW solution and this point is critical with respect to the variations
taken in the same leaf. In our view, the ``averaged'' solution for all
inhomogeneous solutions in a leaf corresponds to the FLRW solution in the
same leaf.

Although our discussions have relied on the known exact solution for the
spherically symmetric case, we have seen that the correspondence between
Eqs.\reff{eq:3-7} and \reff{eq:dV02} is very natural and seems to be
independent of the spherical symmetry. In fact, we have already obtained
a preliminary result that supports a direct generalization of the
present result \cite{MT}. This will appear elsewhere.

We could not discuss a relation to observables like the
distance-redshift relation \cite{SNH}, which will be worth investigating
further.

\section*{Acknowledgments}

The author acknowledges financial support from the Japan Society for the
Promotion of Science and the Ministry of Education, Science and Culture.


\eject

\begin{references}
\bibitem{note:closed} Following the terminology in geometry we use the
  term ``closed'' in the sense ``compact without boundary''.
\bibitem{note:1} However, we cannot have an inhomogeneous solution which
  is completely singularity-free at the interior between the big bang
  and the big crunch. The condition \reff{eq:c2} will in general be
  satisfied only at a finite interval of time from the big bang.
\bibitem{note:4} We assume that the matter flow is irrotational and the
  time-slicing is synchronous.
\bibitem{note:omit} As we have found the integral $\int_0^\pi (\tan
  R\delta f)'\d R$ vanishes, we can check the same for the integral $
  \int_0^\pi \paren{\tan R\delta^2 f+\frac{(\delta f)^2}{2\cos^3 R\sin
      R}}'\d R. $
\end{references}
\end{document}